\documentclass[twocolumn]{aastex631}

\usepackage{physics}
\usepackage{multirow}
\usepackage{soul}

\begin{document}

\title{PIC simulations of nonrelativistic high-Mach-number\\ oblique shocks propagating in a turbulent medium}

\correspondingauthor{Karol Fulat}
\email{fulat@wisc.edu}

\author[0000-0001-6002-6091]{Karol Fulat}
\affil{Department of Astronomy, University of Wisconsin-Madison, Madison, WI 53706, USA}

\author[0000-0003-0404-4943]{Eloise Moore}
\affil{Institute of Physics and Astronomy, University of Potsdam, D-14476 Potsdam, Germany}

\author[0009-0008-0835-2795]{Mahmoud Alawashra}
\affil{Deutsches Elektronen-Synchrotron DESY, Platanenallee 6, 15738 Zeuthen, Germany}

\author[0000-0003-3417-1425]{Michelle Tsirou}
\affil{Institute of Physics of the Czech Academy of Sciences, Prague, 182 00, Czechia}

\author[0000-0002-5680-0766]{Artem Bohdan}
\affiliation{Max-Planck-Institut für Plasmaphysik, Boltzmannstr. 2, DE-85748 Garching, Germany}

\author[0000-0002-2140-6961]{Takanobu Amano}
\affil{Department of Earth and Planetary Science, The University of Tokyo, Tokyo, 113-0033, Japan}

\author[0000-0001-7861-1707]{Martin Pohl}
\affil{Institute of Physics and Astronomy, University of Potsdam, D-14476 Potsdam, Germany}
\affil{Deutsches Elektronen-Synchrotron DESY, Platanenallee 6, 15738 Zeuthen, Germany}

\begin{abstract}
Collisionless shocks are common in astrophysical systems and stand as sites of particle acceleration. While particles at perpendicular shocks may not return to the upstream region, at oblique shocks
a fraction of energetic electrons manage to escape the shock and travel upstream.
An extended region known as the electron foreshock is formed, where these reflected particles drive various instabilities that may promote  
electron acceleration. Here we present the first 
2D3V particle-in-cell (PIC) simulations of electron-ion non-relativistic oblique shocks that explore the interaction of the foreshock with pre-existing compressive turbulence with relative amplitude of 15\% based on interstellar medium estimates.  
We find that pre-existing turbulence influences the emergence and behavior of the whistler-wave instability, as it enhances the amplitudes of the magnetic-field fluctuations and leads to 
larger nonlinear structures. This 
impacts the dynamics of the reflected electrons, resulting in a shorter and hotter electron foreshock. At the end of our simulations, with
pre-existing upstream turbulence we observe 
non-thermal electrons that are more numerous, reach higher energies, and carry 
a larger portion of the total 
energy.
\end{abstract}


\section{Introduction} \label{sec:intro}

Collisionless shocks are mediated by wave-particle interactions instead of binary Coulomb collisions and are known to produce relativistic particles. Diffusive shock acceleration \citep[DSA;][]{Axford1977,Krymskii1977,Bell1978,Blandford1978} provides the basic description of particle acceleration, however it does not take into consideration pre-existing turbulence, which is a common feature in various astrophysical environments. Recent studies of weak heliospheric shocks \citep{Guo2021,Perri2022,Trotta2021,Trotta2023} and shocks propagating in relativistic pair plasma \citep{Tomita2019,Demidem2023,Bresci2023} have shown that pre-existing turbulence does indeed play a significant role in the shock dynamics and particle acceleration.

In this work we focus on nonrelativistic high-Mach-number shocks with sonic and Alfv{\'e}nic Mach numbers \text{$M_\text{s},M_\text{A} \gtrsim 20$}, that are observed in supernova remnants \citep[SNRs; see e.g.,][]{Reynolds2008,Raymond2023} and in the solar system as the bow shocks of Jupiter \citep{Slavin1985}, Saturn \citep{Sulaiman2015,Sulaiman2016}, and Uranus \citep{Bagenal1987}, and occasionally Earth \citep{Sundberg2017,Madanian2021}. 
SNRs are of our primary interest in this study, since they are persistent and resolvable sources that are widely accepted as plausible candidates for the production of galactic cosmic rays. When the supernova ejecta interacts with the ambient medium, strong shocks are formed. Their propagation velocities are typically in the range of \text{$v_\text{sh} \approx 300-10,000~\text{ km s}^{-1}$} \citep{Wang2009,Raymond2023}. Synchrotron radiation produced by accelerated electrons explains radio observations \citep[e.g.,][]{Shklovsky1954,Dubner2015} and nonthermal X-ray emission \citep[e.g.,][]{Koyama1995,Vink2012}. Moreover, $\gamma$ rays emitted from these sources may have a leptonic origin, and be associated with energetic electrons \citep{1996A&A...307L..57P,Brose2021}.

DSA operates only with particles that are significantly more energetic than thermal particles. Thermal electrons, whose characteristic dynamical scales are much smaller than the shock width, cannot directly participate in DSA and must undergo pre-acceleration to reach the required energy threshold.
This is known as the electron injection problem, which is a long-standing challenge in shock-acceleration physics \citep[see, e.g.,][]{Amano2022,Bohdan2023} and a relevant parameter for global models of cosmic-ray sources 
\citep[see, e.g.,][for a review]{Orlando2021}.

Particle-in-cell (PIC) simulations describe collisionless plasmas from first principles and are therefore an excellent tool for investigating electron acceleration and nonlinear processes at electron kinetic scales. They are widely used to study particle acceleration in astrophysical plasmas \citep[see][for a review]{Pohl2020}, especially particle acceleration at SNR shocks \citep{Bohdan2023}. 
In our previous work, we performed for the first time PIC simulations of nonrelativistic high-Mach-number shocks with a pre-existing turbulent upstream medium \citep{Fulat2023}.
As a PIC simulation can only cover fluctuations on scales up to the grid size, the turbulence in the simulation covers about a decade in wavenumber and is restricted to the dissipation range, as opposed to driving on large scales and cascading to small scales \citep{1997ApJ...485..680G}. The latter case requires a fluid approach, and corresponding magnetohydrodynamics (MHD) studies demonstrated that turbulence can distort the shock structure and amplifies the magnetic field via turbulent dynamo \citep[e.g.][]{Inoue2013,Mizuno2014,Ji2016,Hu2022}. Our method of continuous turbulent plasma injection allows to perform simulations for previously unattainable spatial and temporal scales. The results indicated that at perpendicular shocks propagating in compressive turbulence, the behavior of electrons is not significantly different in comparison to the homogeneous upstream medium. Here, we study the shock-turbulence interplay at oblique nonrelativistic high-Mach-number shocks, where shock-reflected electrons drive various instabilities in the electron foreshock. 
For homogeneous upstream media \citet{Bohdan2022} found that in the outer foreshock region electrostatic electron-acoustic waves are present, while the inner foreshock is dominated by electromagnetic oblique whistler waves. The electron-acoustic waves carry less energy and can affect only less than 0.1\% of the incoming electrons \citep{Morris2022}. The whistlers are gyroresonant with the shock-reflected electrons and have a significant effect on particle dynamics. In their 2D PIC simulations \cite{Morris2023} found a significant fraction of nonthermal electrons in the downstream energy spectrum, consistent with previous 1D simulations \citep{Xu2020,Kumar2021}. Electrons are efficiently accelerated by a mechanism similar to stochastic shock drift acceleration (SSDA), for the first time observed in 3D shock simulations by \cite{Matsumoto2017} {and described in detail by \citet{Katou2019,Amano2022}.} \cite{Morris2023} observed more efficient acceleration than predicted by SSDA, possibly due to turbulent electric fields driven in the nonlinear regime of the whistler instability.

This paper explores the structure of oblique shocks, and the electron acceleration efficiency at them, for an upstream medium with pre-existing compressive turbulence.
This paper begins with a description of the setup of simulations used in this study (Section 2) followed by a presentation of the results in Section 3. Finally, a summary and discussion are given in Section 4.

\section{Simulation setup} \label{sec:sim_setup}

In this work we use the fully relativistic PIC code THATMPI \citep{Niemiec2008}, 
which tracks two spatial and all three velocity components of the macro-particles describing the plasma (2D3V configuration). This setup optimizes computational resources and allows to run simulations with higher resolution compared to full 3D simulations while maintaining a reasonably accurate description of shock physics.

The simulation setup is similar to that of \cite{Fulat2023}, in which we investigated perpendicular shocks with turbulent and homogeneous upstream media. Here, we briefly outline the setup and highlight differences. The shock wave in our simulation is established by reflecting a plasma beam off a conducting wall. The interaction between the incoming and the reflected beams then creates a shock. Since the simulation frame is equivalent to the downstream rest frame, the upstream plasma must be continuously replenished.

Whereas in previous simulations with a homogeneous upstream medium the upstream plasma was added in thin layers, plasma injection incorporating synthetic turbulence requires a new method \citep{Fulat2023}. Pre-fabricated slabs of turbulent plasma are simulated in square boxes with periodic boundary conditions that are as wide as the main simulation box.  In these we impose quasi-isotropic bulk-velocity fluctuations on scales of a few ion inertial lengths. Following the plasma response over a few ion Larmor times we see that the initial fluctuations evolve into correlated variations of the electromagnetic fields, density and bulk velocity of the plasma. Over time, the turbulence decays, and the energy is transferred to heating, which for a given intended sonic Mach number restricts the turbulence amplitude to about 15\% for the density fluctuations and around 10\% for the magnetic field, $\delta B/B_0$. When the turbulence is established, we inject these slabs into the far-upstream region of the shock simulation, after matching them at the interface to prevent numerical transients. This matching technique provides a smooth transition of current densities and electromagnetic field between turbulent plasma over a selected region, which is obtained by interpolation of the field and the particle distribution function. In the shock simulation, the turbulence evolves further until the plasma passes through the shock. Additional technical details are found in section 2 and in the appendix of \citet{Fulat2023}.

Oblique-shock simulations require extended boxes in order to contain the electron foreshock. For pre-existing turbulence injected at the outer right boundary, advection towards the shock takes longer than in perpendicular-shock simulations. Therefore, the turbulence decay time must be increased, which motivates the use of larger simulation boxes than in our previous study, \text{$L_y \approx 12 \lambda_\text{si}$} instead of \text{$L_y = 6 \lambda_\text{si}$}. A secondary benefit of this choice is that nonlinear structures driven by the whistler instability with sizes up to $5\lambda_\text{si}$ can be fully captured \citep[see][]{Morris2023}.

We performed two simulations, one with density fluctuations of amplitude \text{$\var{n}/n_0=15\%$} (run~$\text{T}$) and one with a homogeneous upstream medium (run $\text{H}$). The fluctuation amplitude, \text{$\delta n$}, is calculated as the root mean square (rms) of the particle density measured in small tiles that are a quarter of ion skin length in size, and \text{$n_0$} is the mean density. The magnitude of the density fluctuations on kinetic scales for SNRs are unknown. In the simulation, larger fluctuation levels would lead to strong heating that compromises the sonic Mach number. Our choice of amplitude avoids that and corresponds to measurements in the heliosphere \citep{Carbone2021} and the local interstellar medium \citep{Lee2020,Ocker2021,Fraternale2022}, which show \text{$\var{n}/n_0 \lesssim 10\%$}.

\begin{table}[t!]
    \begin{center}
        \caption{Relevant plasma parameters for the initial setup of the simulations used in this study.}
        \begin{tabular}{lc|c}
            \hline
            \hline
            \multicolumn{2}{c}{Parameter} & Value \\
            \hline
            sonic Mach number & $M_\text{s}$ & 36 \\
            Alfv{\'e}nic Mach number & $M_A$ & 32 \\
            plasma beta & $\beta_p$ & 1 \\
            shock speed & $v_\text{sh}$ & $0.264c$ \\
            electron thermal velocity & $v_{\text{th},e}$ & $0.057c$ \\ 
            number of particles per cell & $n_0$ & 20 \\
            electron skin depth & $\lambda_\text{se}$ & $20\Delta$ \\
            mass ratio & $m_i/m_e$ & 100 \\
            \hline
            \multirow{2}{1in}{turbulence level} & \multirow{2}{0.35in}{$\delta n/n$} & 15\% (run T) \\
            & & \phantom{5}-\phantom{\%} (run H) \\
                        \hline
        \end{tabular}
        \label{tab:sim_params}
    \end{center}
    \footnotesize{Here, $\Delta$ represents the computational cell size, and $c$ denotes the speed of light. Refer to the main text for definitions of all parameters.}
\end{table}

Table~\ref{tab:sim_params} shows the relevant 
plasma parameters for each simulation. The initial large-scale magnetic field has an out-of-plane configuration, {$\vb{B}_0=B_0(\cos\theta_0, 0, \sin\theta_0)$}, where the obliquity angle is \text{$\theta_0 = 60^\circ$}. This choice provides the same number of degrees of freedom and adiabatic index of plasma as a 3D setup. It was also used in \cite{Morris2023}, since it fully captures the oblique whistler waves. 

We initialize the upstream plasma with twenty particles per cell for both ions and electrons, \text{$n_0=n_i=n_e=20$}, where the subscripts ``$i$'' and ``$e$'' represent ions and electrons respectively. The plasma flows toward the reflecting wall with velocity \text{$\mathbf{v_0}=v_0\hat{x}$}, where \text{$v_0=-0.2c$} and $c$ is the speed of light. 
The expected shock speed in the upstream frame is \text{$v_\mathrm{sh} \simeq 0.264c$}.

In the shock simulation the particles are initially in thermal equilibrium at {$T_e \approx T_i \approx 1.6\cdot10^{-3}m_ec^2/k_B\approx 1.6\cdot10^{-5}m_ic^2/k_B\approx 10^7 \text{ K}$}, where $T$ are temperatures, $m$ particle masses, and \text{$k_B$} is the Boltzmann constant. For such non-relativistic conditions the thermal speed of electrons is $v_{th,e} = \sqrt{2k_BT_e/m_e}\simeq 0.057c$, and ten times smaller for ions, on account of the ion-to-electron mass ratio $m_i/m_e=100$. The plasma beta, defined as the ratio of the thermal pressure to the magnetic pressure, is $\beta_p \approx 1$ for all runs. The sonic Mach number then follows as $M_\text{s}=v_\mathrm{sh}/c_S\approx 36$ for all runs, where $c_S=\sqrt{\Gamma k_B(T_e+T_i)/m_i}\approx 0.0074c$, where $c_S$ is the sound speed and $\Gamma=5/3$ is the adiabatic index. Likewise, the Alfv{\'e}n speed $v_A=B_0c/\sqrt{n_0(m_e+m_i)}=0.00829c$ leads to an  Alfv{\'e}nic Mach number $M_A \approx 32$. The expected compression ratio at the shock is $r \simeq 3.97$. 

The time step, \text{$\delta t=1/40\,\omega_{pe}^{-1}$}, scales with the electron plasma frequency, \text{$\omega_{pe}=\sqrt{e^2n_e/(\epsilon_0m_e)}$}, where e is the elementary charge and \text{$\epsilon_0$} is the vacuum permittivity. We evolve the system for approximately \text{$20\,\Omega_{i}^{-1}$} in ion cyclotron times,  $\Omega_i^{-1}=m_i/(e B_0) \approx 48,000 \delta t$. The electron skin depth is $\lambda_\text{se}=c/\omega_{pe}=20 \Delta$, where $\Delta$ represents the cell size; for ions it is higher by a factor $\sqrt{m_i/m_e}=10$.

Our choice of simulation parameters is a compromise between computational expedience and the accurate modeling of the shocks. By using a shock speed that is higher than that typically found in SNRs, but still nonrelativistic, we are able to extend the duration of our simulations. In addition, these parameter choices allow for a direct comparison with previous studies of high-Mach-number shocks.

\begin{figure*}[t!]
    \centering
    \includegraphics[width=0.95\linewidth]{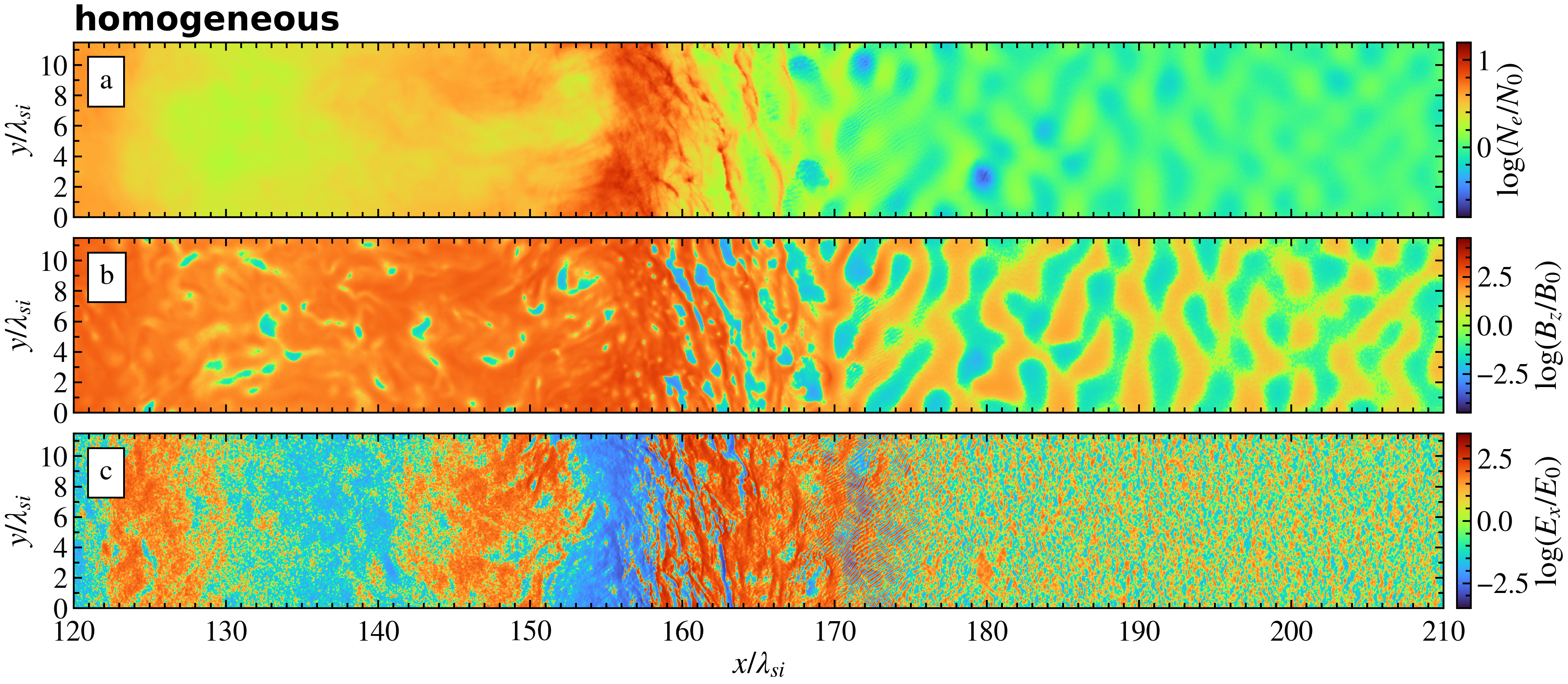}
    \includegraphics[width=0.95\linewidth]{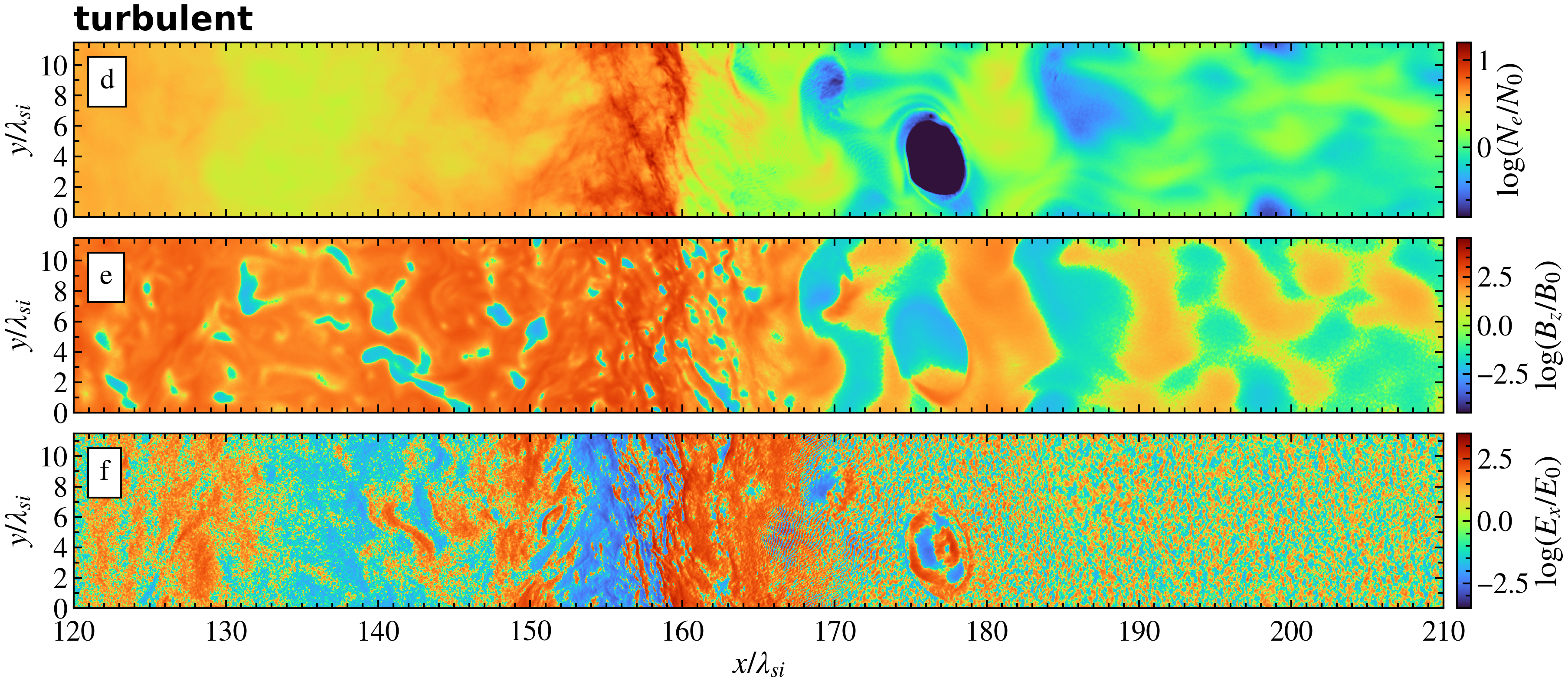}
    \caption{Maps of the electron number density (panels a and d), the $z$-component of the magnetic (panels b and e), and the $x$-component of electric fields (panels c and f) at an oblique shock with pre-existing upstream density fluctuations (run $\text{T}$, panels d,e, and f) and with a homogeneous upstream medium (run $\text{H}$, panels a,b, and c), both at $t\approx 18\,\Omega_i^{-1}$. For the $B_z$ map, the initial upstream field was subtracted. The scaling of the fields is logarithmic and sign preserving, e.g., for $B_x/B_0$ it is $\text{sgn}(B_x)\cdot[2+\log \{\text{max}(10^{-2},|B_x|/B_0)\}]$. }
    \label{fig:oblique_maps}
\end{figure*}

\section{Results} \label{sec:results}
In this section, we compare the results of simulations with an initially turbulent (run~$\text{T}$) and a homogeneous (run~$\text{H}$) upstream medium. Our focus is on the influence of pre-existing turbulence on the global shock structure, the shock velocity, magnetic-field amplification, and the properties of the nonlinear structures developed by the oblique whistler instability. We then investigate how the plasma waves are affected by the pre-existing turbulence. Moreover, we examine the influence of the turbulence on the shock-reflected electron distribution. Finally, we study the electron energy spectra in the downstream region. 

\subsection{Global shock structure}
Generally, shock-reflected ions determine the structure of supercritical shocks. However, at oblique shocks, a significant portion of the reflected particles are electrons. They form a complex structure known as the electron foreshock. Of the two main instabilities driven there, the electron-acoustic waves are expected to be weak for the out-of-plane configuration of our simulation (\cite{Bohdan2022}, \cite{Morris2022}), and they are not considered in the present analysis. After a few ion Larmor times, the electromagnetic oblique-whistler instability driven in the inner foreshock develops into its nonlinear stage and creates wavepackets that are advected with the incoming flow. They interact with the shock and modify the shock foot and ramp.

\begin{figure*}[t!]
    \centering
    \includegraphics[width=0.95\linewidth]{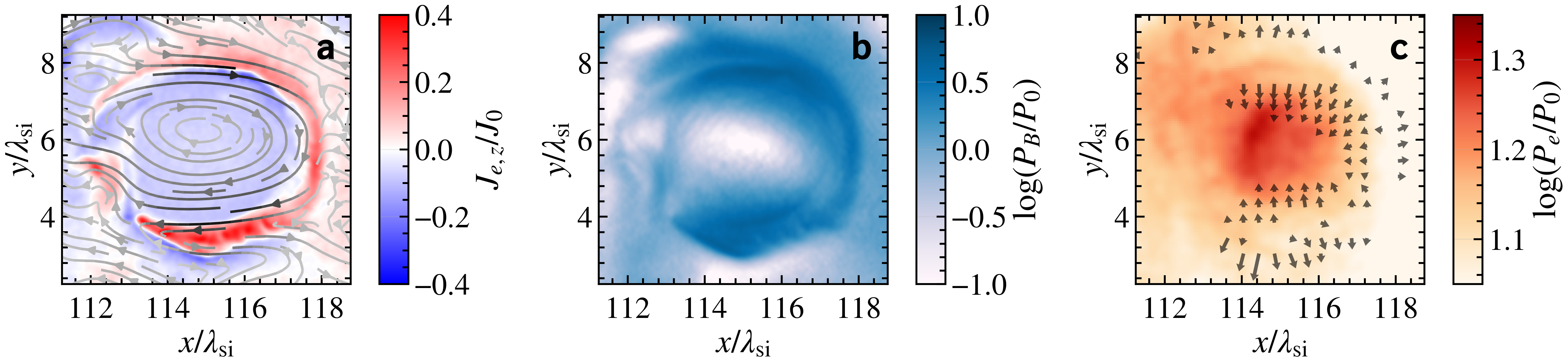}
    \caption{Structure of an example nonlinear cavity observed in the foreshock of the turbulent simulation. Panel a shows the $J_z$ component of the electron current density normalized to $J_0=q_en_0v_0$ with superimposed magnetic-field lines, their colour-shade representing their strength. Panels b and c show the magnetic and electron pressures, respectively, that are normalized by the initial pressure, $P_0=B_0^2/2\mu_0+n_0k_B(T_e+T_i)$. Arrows on panel c represent the $\vb{J}\times\vb{B}$ force field.
    }
    \label{fig:cavity}
\end{figure*}

\begin{figure*}[t!]
    \centering
    \includegraphics[width=0.49\linewidth]{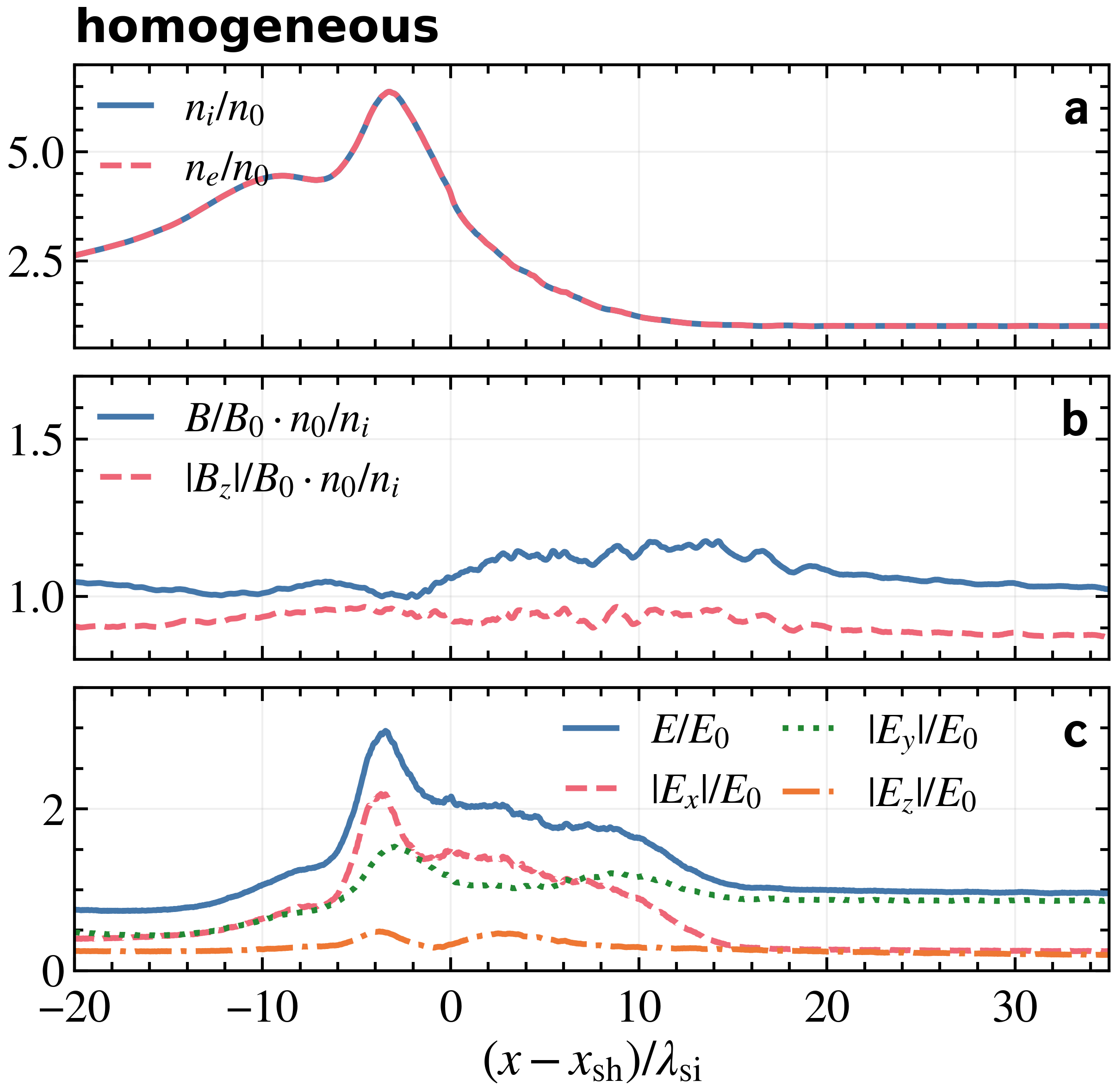}
    \hfill
    \includegraphics[width=0.476\linewidth]{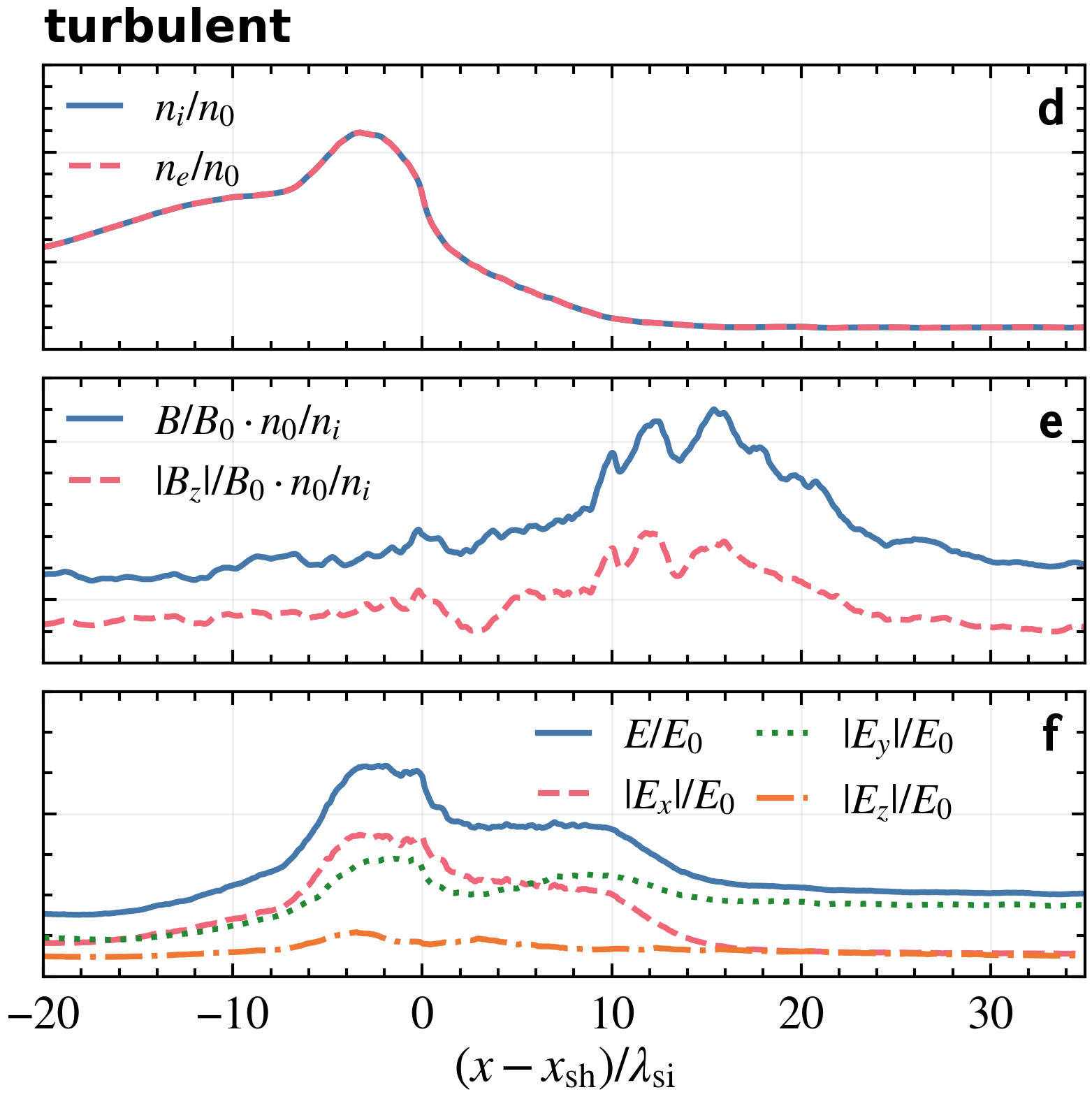}
    \caption{The ion and electron number density profile (panels a,d), the ratio of the magnetic field strength to the density (panels b,e), and the electric field profiles (panels c,f) averaged over the $y$-direction, for run $\text{H}$ (panels a,b, and c) and $\text{T}$ (panels d,e, and f). The profiles are time-averaged from $t\Omega_i\approx7$ to $t\Omega_i\approx20$. 
    }
    \label{fig:oblique_prof}
\end{figure*}

Figure~\ref{fig:oblique_maps} compares the electron number density, the $z$-component of the magnetic field, and the $x$-component of the electric field at a late stage of the two simulations, at $t\approx 18\,\Omega_i^{-1}$. 
The plot covers only the inner part of the electron foreshock, where the whistler waves are present. Panels (a) and (d) of Figure~\ref{fig:oblique_maps} show that these waves are stronger closer to the shock front since the energy density of the reflected electrons decreases with distance from the shock. The nonlinear structures developed by the instability, such as cavities with bipolar magnetic-field features, appear at about \text{$t \approx 9\Omega_i^{-1}$} for the homogeneous simulation and at about $t \approx 7\Omega_i^{-1}$ in the presence of pre-existing turbulence. For the initially turbulent upstream medium in run $\text{T}$ they can be as large as 7$\lambda_\text{si}$, to be compared to roughly 2$\lambda_\text{si}$ for run $\text{H}$. 

Figure~\ref{fig:cavity} shows the structure of a representative cavity from the turbulent simulation, with all quantities computed in the upstream rest frame. Since the bipolar magnetic field has mostly in-plane components, the magnetic-field lines depicted in panel (a) thus give a fair impression of the full magnetic-field structure. This field is generated by the $J_z$ component of the electron current density. The negative $J_z$ current derives from the drift of reflected electrons along the background magnetic field which is only $30^\circ$ off the $z$ axis. The lighter color of the streamlines in panel (a) indicates that the magnetic field strength decreases towards the cavity center. This is also apparent in panel (b), which presents the magnetic pressure. Panel (c) illustrates a high electron pressure inside the structure, where the hot, low-density reflected electrons are accumulated\footnote{This is verified by comparing the electron momentum distributions inside the cavity and at its boundary.}. The corresponding outward-directed force is compensated by the Lorentz force on these particles that tends to confine them inside the structures
\citep{Morris2023}, which is indicated by the arrows in this panel. At the cavity boundary, the Lorentz force pushes the dense and cold background particles outward.

The average perpendicular electron momentum inside the cavity is approximately $p_\perp/m_ec\approx4$. With the magnetic field roughly twice the background value, the gyroradius is $r_g\approx 2\lambda_\text{si}$, comparable to the structure radius $R\approx2\lambda_\text{si}$. The electrons inside the cavity have a non-zero velocity in the $x$-direction, $v_x\approx0.4c$, and so their cavity-crossing time is $\tau=2R/v_x\approx 10/ \omega_{pi}$. This is larger than the reciprocal of the electron gyrofrequency, $\gamma_\mathrm{e}\Omega_\mathrm{e}^{-1} \approx 2.5/ \omega_{pi}$, for the typical Lorentz factor $\gamma_\mathrm{e}\approx 4$. Therefore, shock-reflected electrons of moderate energy are likely confined within, which is consistent with previous studies \citep[see Section 4.5 in ][]{Morris2023}.
The balance of the magnetic and electron pressures causes these structures to inflate as they are advected towards the shock. Ion pressure is omitted from the discussion, as this species exhibits no distinct pressure signatures in these cavities. The properties of the cavities observed in the homogeneous simulation are analogous, although their scale is smaller and their electron pressure is lower.

In Figure~\ref{fig:oblique_maps} the shock foot is located between {$x =170\lambda_\text{si}$ and $x=175\lambda_\text{si}$.
In its vicinity, the Buneman instability drives strong electrostatic modes which are visible as striped motifs in the \text{$E_x$} maps. Hereafter, we use them as an indicator of the shock position}. We note that the cavities in 
Figure~\ref{fig:cavity} emerge beyond the shock foot in the foreshock region.
The shock ramp contains the characteristic filaments of the Weibel instability that is driven by counter-streaming ions. It is strongest for wavevectors perpendicular to the relative velocity between the shock-reflected and incoming ions.  
The filaments are 
oblique because the bulk-gyrating ions have a non-zero \text{$v_y$} velocity component. 
The transition to the downstream region is marked by the overshoot-undershoot pattern that extends over roughly 5-10$\lambda_\text{si}$ behind the shock front.

\begin{figure}[t!]
    \centering
    \includegraphics[width=0.98\linewidth]{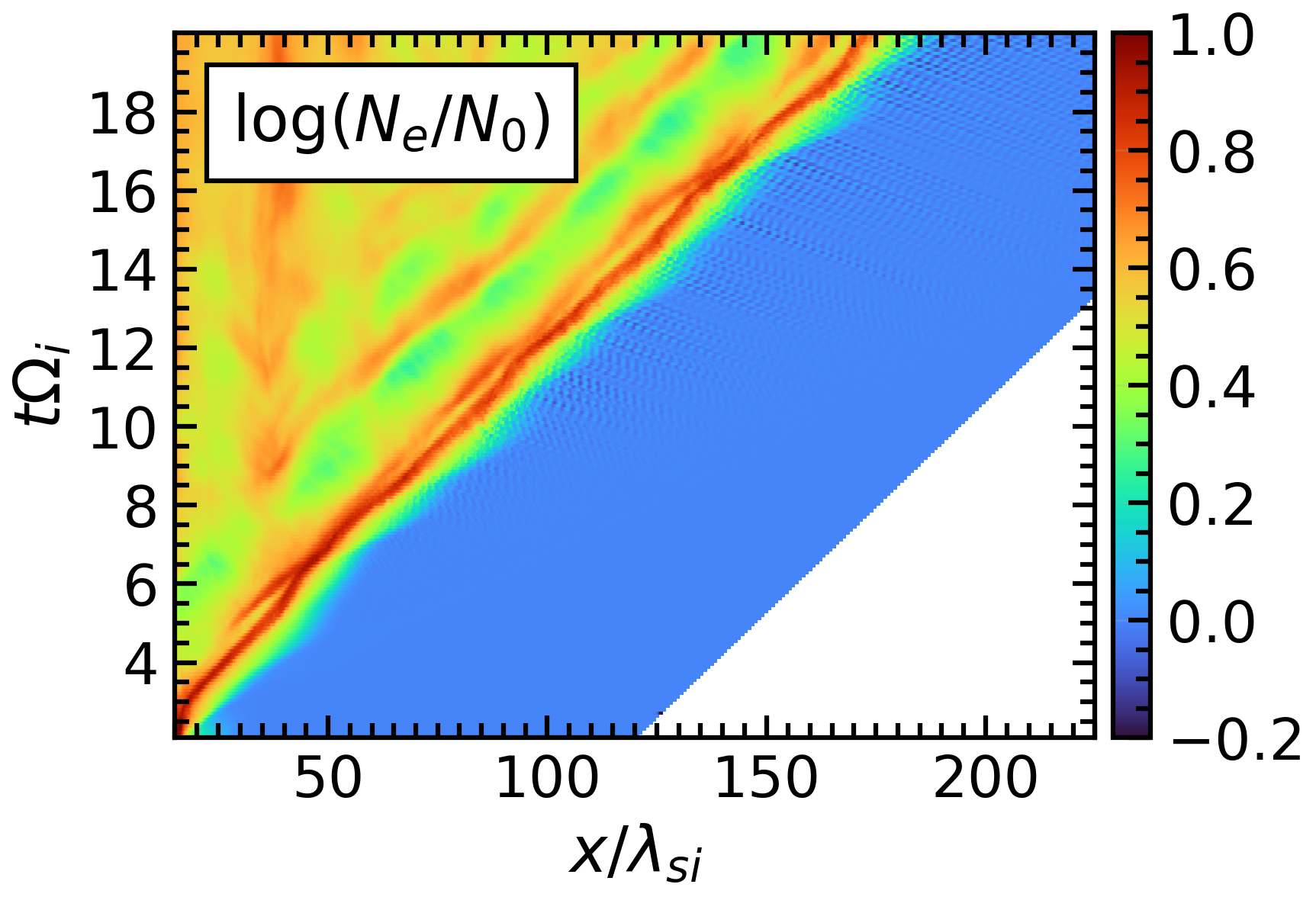}
    \includegraphics[width=0.98\linewidth]{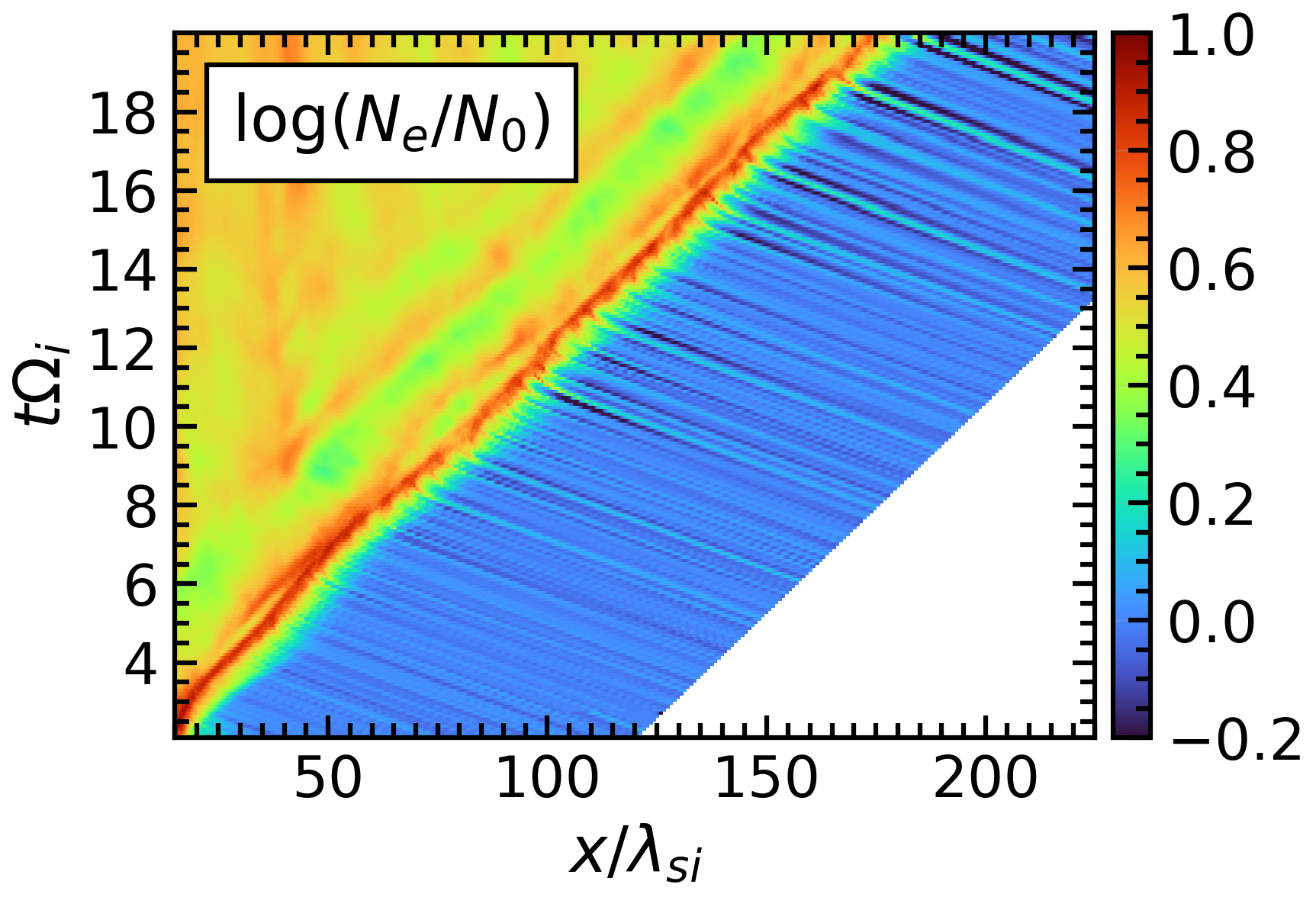}
    \caption{Evolution of the electron number density averaged over the $y$-direction for run $\text{H}$ (top panel) and $\text{T}$ (bottom panel). For visualization purposes, the presentation is truncated $110\lambda_\text{si}$ ahead of the shock.}
    \label{fig:oblique_dens_prof}
\end{figure}

Figure~\ref{fig:oblique_prof} shows profiles averaged over the $y$-direction of the ion number density and the absolute strength of the magnetic and electric fields. The profiles are time-averaged, starting from the beginning of the regular shock reformation at \text{$t\Omega_i\approx7$}, up until the end of the simulation, (see the top panel in Figure~\ref{fig:oblique_dens_prof} and the subsequent paragraph for further details). 
Comparing the density profiles for runs $\text{H}$ and $\text{T}$, shown in panels (a) and (d), reveals that with pre-existing turbulence 
the overshoot is wider in size and the undershoot is less pronounced. The compression ratio is consistent with an MHD shock expectation for both simulations. 
As previously mentioned, the out-of-plane configuration of the external magnetic field weakens the presence of the Weibel instability. Therefore, the growth of the magnetic field in the shock transition region is primarily due to the compression of the \text{$B_z$} component of the initial magnetic field \text{$\vb{B_0}$}, as \text{$|B_z|/n_i$} is nearly constant.
In both runs, the main amplification of the magnetic field begins at about \text{$30\lambda_\text{si}$} ahead of the shock front (Figure~\ref{fig:oblique_prof}b,e) which corresponds to the region where the nonlinear structures develop.
For the run with the pre-existing density fluctuations, the
magnetic field in the corresponding region is stronger by approximately 25\text{$\%$}. The constant \text{$E_y$} component of the electric field in the upstream region is the motional electric field, \text{$\vb{E}_0=-\vb{v}_0\times\vb{B}_0$}. In the shock foot region, the Buneman instability enhances \text{$E_x$} and \text{$E_y$}, as the wavevectors lie in the $x$-$y$ plane. In the ramp region, a charge-separation field is present, indicated by the increase of the \text{$E_x$} in Figure~\ref{fig:oblique_prof}c,f. Both \text{$E_x$} and \text{$E_y$} are amplified in the overshoot region. Though poorly observable, the magnitude of \text{$E_z$} is also enhanced. Strong electric fields at the shock ramp are paramount for efficient electron acceleration. 
\citet{Morris2023} report that the energy gain of electrons at the shock is much larger than the work done by drifting along the motional electric field, with \text{$E_x$} and \text{$E_z$} significantly contributing. This leads to a more efficient acceleration than typically expected from SSDA.

Figure~\ref{fig:oblique_dens_prof} shows the evolution of the $y$-averaged electron density for simulations with homogeneous (run $\text{H}$, top panel) and turbulent upstream plasma (run $\text{T}$, bottom panel). Following a swing-in phase that lasts about $t \approx 7\Omega_i^{-1}$, cyclic shock reformation is visible as modulations of the shock front in run H. The reformation period is, on average, roughly $1.6\Omega_i^{-1}$, which is consistent with the results for perpendicular shocks, except for the delayed appearance \citep[see, e.g.,][]{Wieland2016,Bohdan2017,Fulat2023}. The dark blue stripes located ahead of the shock front in the top panel represent the whistler-driven structures. They become evident at about $t \approx 7 - 9\Omega_i^{-1}$, reflecting the nonlinear stage of the whistler instability. In the bottom panel of Figure~\ref{fig:oblique_dens_prof} representing run T, the $x$-$t$ traces of the pre-existing density fluctuations are visible as straight features at all times. 
The amplitude of the upstream perturbation increases with time, and the features develop sub-structure, suggesting that, when the upstream medium is initially turbulent, the nonlinear structures develop earlier, are stronger, and/or appear more frequently. Shock-front corrugation is superimposed on the reformation pattern for run $\text{T}$. The structure of the overshoot is blurred, and the peak density is slightly lower than for homogeneous conditions.

\subsection{Wave modes in the foreshock region}\label{foreshock_waves}
Previous 
shock simulations with a homogeneous upstream medium demonstrated that, for out-of-plane magnetic-field configurations, oblique whistler waves are the dominant modes in the inner foreshock region \citep{Morris2023}. With a few ion skin lengths \citep[see Fig.7 of][]{Bohdan2022}, their wavelengths are comparable to the characteristic length scales of the pre-existing turbulence in run H. Consequently, it is challenging to distinguish the modes in the power spectrum of magnetic-field fluctuations. Instead, the right-hand circular polarization of the whistler waves can be used for identification.

In our simulations, the wavevector of a whistler wave must lie in the $x$-$y$ plane, \text{$\vb{k}=(k_x,k_y)$}. Defining \text{$\theta(\vb{k}) \equiv \angle (\vb{k},\vu{x}) = \arctan(k_y/k_x)$}, the Fourier amplitudes of waves for right-hand circular (subscript ${R}$) or left-hand circular (subscript ${L}$) polarization, $B_R$ and $B_L$, can be written as 
\begin{align}
    \widetilde{B_R}(\vb{k}) = \widetilde{B_z}(\vb{k}) &- i \sin[\theta(\vb{k})] \widetilde{B_x}(\vb{k}) \label{eq:FT_B_R} \nonumber \\ &+ i \cos[\theta(\vb{k})] \widetilde{B_y}(\vb{k}), \\
    \widetilde{B_L}(\vb{k}) = \widetilde{B_z}(\vb{k}) &+ i \sin[\theta(\vb{k})] \widetilde{B_x}(\vb{k})  \label{eq:FT_B_L} \nonumber \\ &- i \cos[\theta(\vb{k})] \widetilde{B_y}(\vb{k}),
\end{align}
where the tilde represents the transform. A derivation of this expression can be found in Appendix~\ref{sec:A1}. These quantities are calculated in the upstream reference frame to preserve the intrinsic polarization pattern of the waves.
Calculating the absolute values yields power spectra characteristic of right-hand and left-hand circularly polarized modes, and Parseval's theorem directly provides the energy density of the magnetic field associated with the modes.

\begin{figure}[t!]
    \centering
    \includegraphics[width=0.98\linewidth]{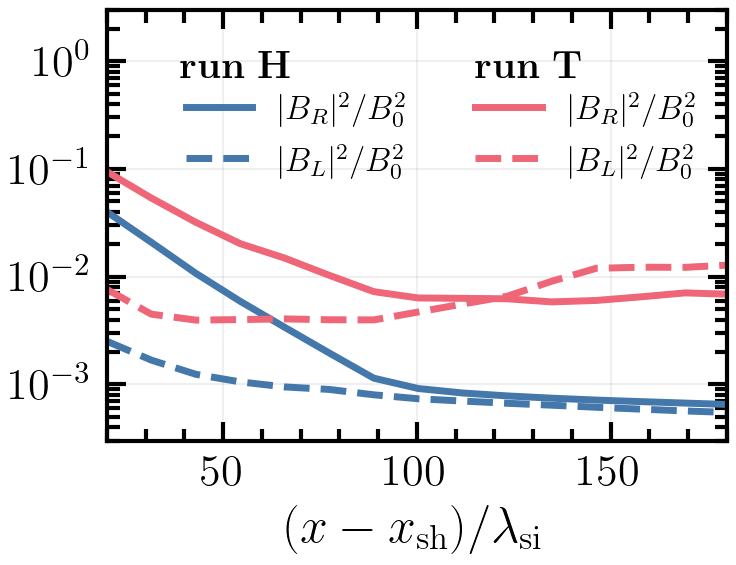}
    \caption{The energy density of the right (solid lines) and left (dashed lines) circularly polarized modes in the foreshock region for run H (blue lines) and run T (red lines). The energy density is normalized to the energy density of the initial magnetic field.}
    \label{fig:polar_shock}
\end{figure}

Figure~\ref{fig:polar_shock} shows the energy density of modes with right or left-hand circular polarization in the inner foreshock region for both simulations. The energy densities at a given location are determined from the power spectra of \text{$B_R$} and \text{$B_L$} calculated within a square region of size: \text{$x \in [x-L_y/2,x+L_y/2], y \in [0,L_y]$}, time-averaged over the reformation cycle that follows the emergence of the nonlinear structures in the homogeneous simulation, \text{$11 \lesssim \Omega_i t \lesssim 12.5$}, and 
normalized to the energy density of the background magnetic field. With a homogeneous upstream medium (run H), \text{$B_R$} and \text{$B_L$} have small amplitudes at distances $x-x_\text{sh} \gtrsim 90 \lambda_\text{si}$, indicating that there are no strong electromagnetic waves this far upstream.
In the range \text{$20 \lambda_\text{si}\lesssim x-x_\text{sh} \lesssim 90 \lambda_\text{si}$}, the magnitude of \text{$B_R$} grows exponentially, indicating the driving of right-hand polarized waves. At \text{$x-x_\text{sh} \approx 20$} the nonlinear structures start to emerge.

Due to their relatively low phase and group velocities, whistler waves comove with the background plasma. Consequently, the spatial amplification of their amplitudes along the $x$-direction is equivalent to temporal growth, as \text{$( x-x_\text{sh})= C-v_\mathrm{sh}t$}. The
exponential slope in Figure~\ref{fig:polar_shock} hence is a measure of the growth rate of the right-hand polarized mode, $\gamma$, as
\begin{equation}
    |B_R|^2 \propto \exp\left(-2\gamma  \frac{ x-x_\text{sh}}{v_\mathrm{sh}}\right).
    \label{eq:growth_rate}
\end{equation}
The fitted growth rate equals 
$\gamma_\text{H}=(8.16 \pm 0.33)\times 10^{-3} \Omega_e$ for the homogeneous run, where the uncertainty range is for 95\text{$\%$} confidence level (\text{$2\sigma$}). \citet{Bohdan2022} conducted periodic-box simulations with particle distributions close to those in the electron foreshock of our shock simulations, and they found consistent agreement with the results yielded by linear analysis of oblique whistler waves. 

For run T, the energy densities for the two polarizations are comparable in the far-upstream region and reflect the properties of the pre-defined compressive turbulence 
(see Appendix~\ref{sec:A2} for further details). As in the homogeneous simulation, at around \text{$x-x_\text{sh} \approx 90 \lambda_\text{si}$}, \text{$B_R$} starts to grow exponentially toward the shock front. An exponential-function fit yields a growth rate of $\gamma_\text{T}=(5.8 \pm 0.58)\times 10^{-3} \Omega_e$,
which is lower than that derived from run H. This may be a consequence of the higher temperature of the shock-reflected electrons with pre-existing turbulence (see Section~\ref{foreshock_properties}), that leads to lower growth rates \citep[see Figure 11 in][]{Bohdan2022}. Linear analysis would be only partially applicable, as it would ignore the pre-existing fluctuations.

The amplitude of the whistler waves at $x-x_\text{sh} \approx 20 \lambda_\text{si}$ remains constant for the duration of our simulations, $|B_R|^2/B_0^2 \simeq 0.04$ and $|B_R|^2/B_0^2 \simeq 0.1$, but in both runs the wave growth begins at progressively larger distances from the shock. 

\begin{figure*}[t!]
    \centering
    \includegraphics[width=0.49\linewidth]{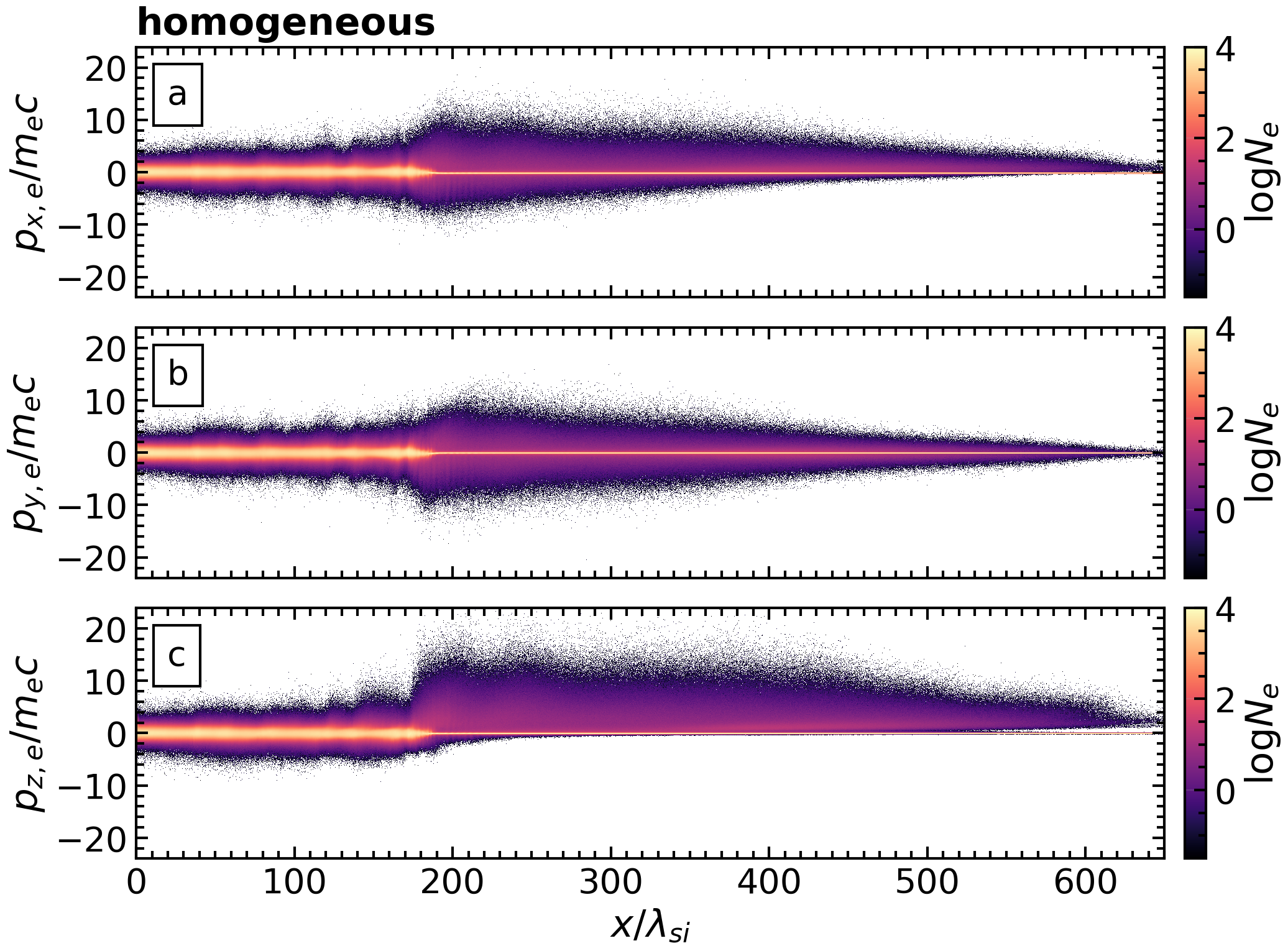}
    \includegraphics[width=0.49\linewidth]{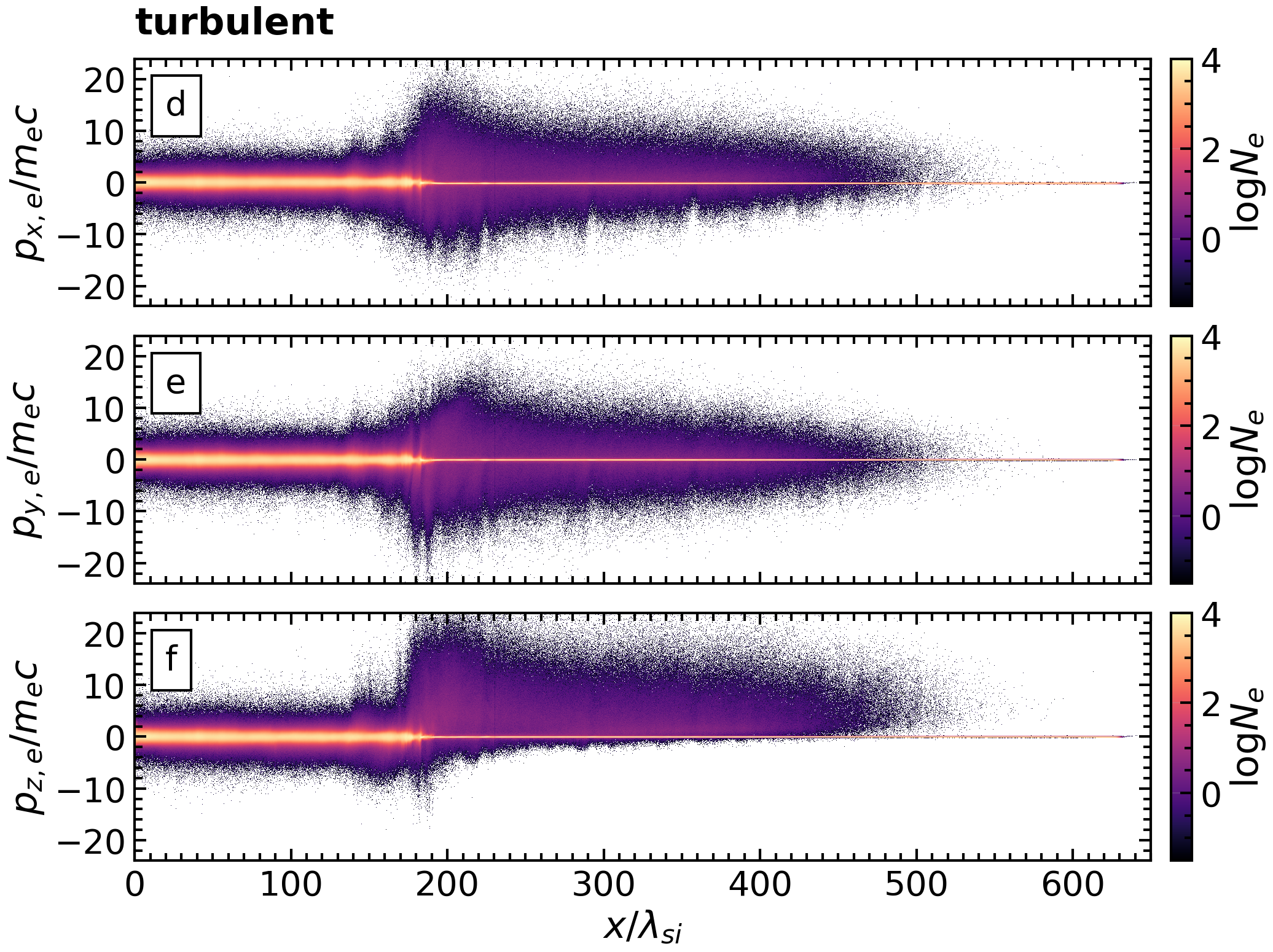}
    \caption{Electron phase-space distributions for run $\text{H}$ (panels a,b, and c) and run $\text{T}$ panels (d,e, and f) at time $t \approx 20 \Omega_i^{-1}$.}
    \label{fig:oblique_phase}
\end{figure*}

\subsection{Electron foreshock properties}\label{foreshock_properties}
Figure~\ref{fig:oblique_phase} compares the electron phase-space distributions for runs $\text{H}$  and $\text{T}$ at the end of the simulations (at about 20 ion Larmor times), when the shock front is located at $x_\text{sh} \approx 177\lambda_\text{si}$. Ahead of the shock, the cold incoming particles are represented by a narrow strip centered around $(p_x,p_y,p_z)/mc=(-0.2,0,0)$. The other particles present at $x \gtrsim x_\text{sh}$ are the shock-reflected electrons.

When the upstream medium is initially homogeneous (panels a,b, and c), the electron foreshock extends up to $670\lambda_\text{si}$. The momentum spread of the shock-reflected electrons decreases with distance from the shock, and so does the amplitude of the whistler waves with which the electrons are gyroresonant, suggesting a declining 
scattering efficiency 
(see Section \ref{foreshock_waves}). 
In contrast, for turbulent upstream conditions, run $\text{T}$ (panels d, e, and f), the electrons do not travel that far, and the right boundary of the foreshock lies at $540\lambda_\text{si}$, closer to the shock by 25\% or roughly five Larmor radii of the upstream ions. Also, the distribution of the reflected electrons appears wider, hence hotter, suggesting enhanced scattering in the foreshock. 

\begin{figure}[t!]
    \centering
    \includegraphics[width=1.0\linewidth]{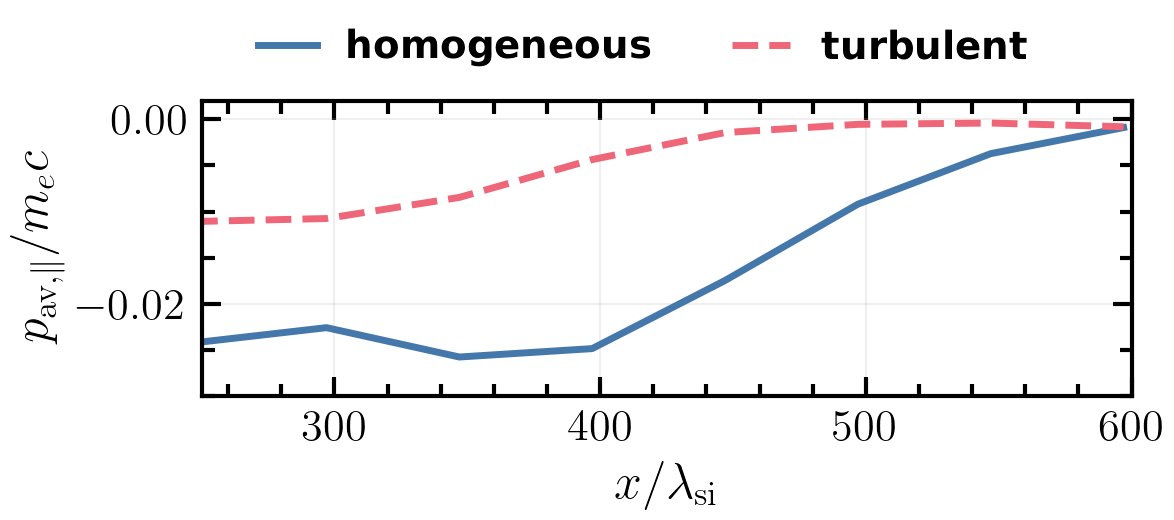}
    \caption{The parallel component of the average momentum of the background electrons in the upstream rest frame.}
    \label{fig:momentum_profile_bkg}
\end{figure}

\begin{figure}[t!]
    \centering
    \includegraphics[width=1.0\linewidth]{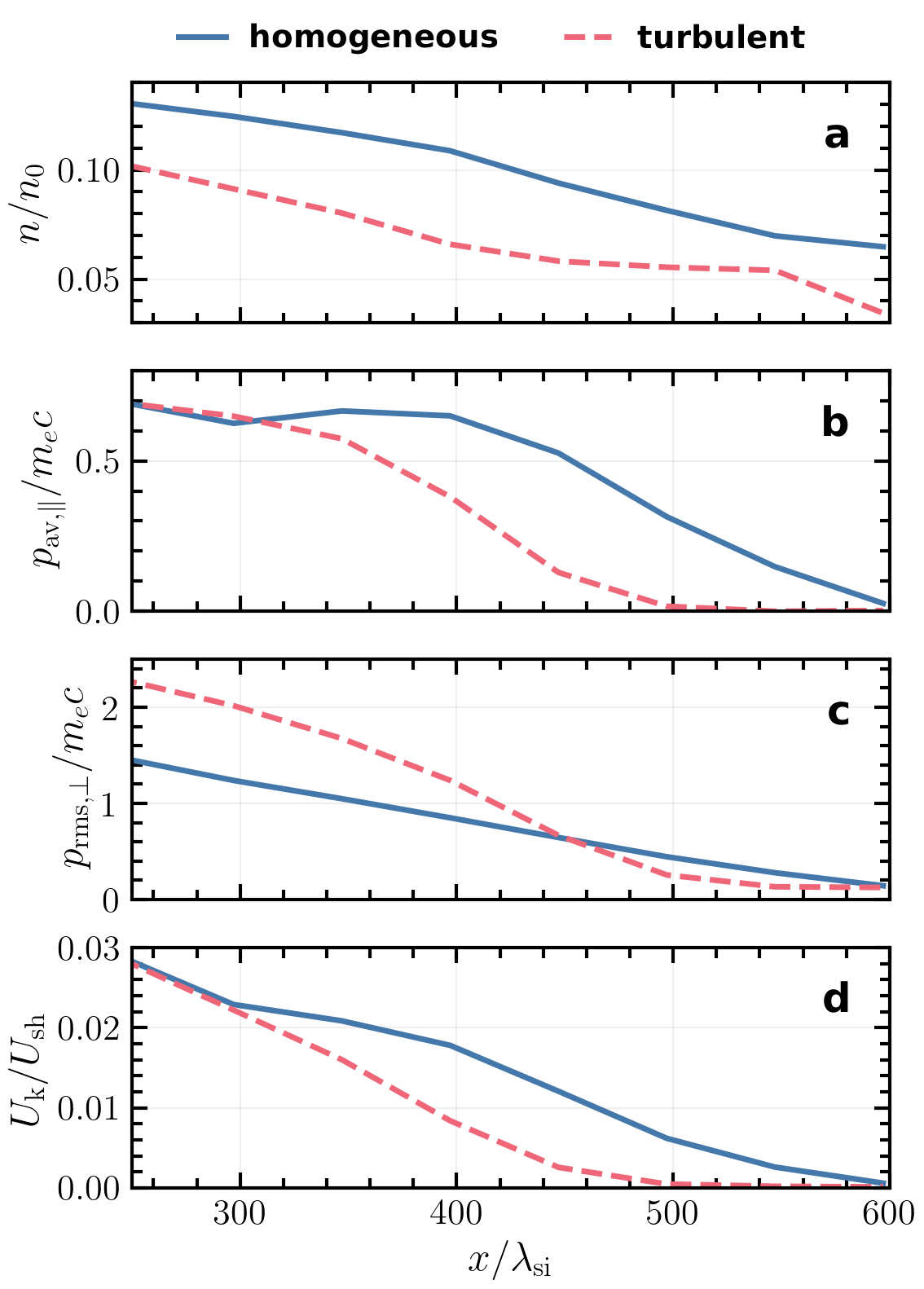}
    \caption{The number density (panel a), the parallel component of the average momentum (panel b), the perpendicular component of the rms momentum (panel c), and the energy density (panel d) of the shock-reflected electrons. All quantities are calculated in the upstream rest frame. }
    \label{fig:momentum_profile_ref}
\end{figure}

To study the properties of shock-reflected electrons in greater detail, we separate them from the incoming electron beam. We compute the three-dimensional momentum distributions of electrons in the upstream rest frame within regions of $50\lambda_\text{si}$ width and height $L_y$, choosing a coordinate system in which one axis is aligned with the initial magnetic field. Then, we fit a drifting non-relativistic Maxwellian to these distributions, assuming it represents the background (incoming) electrons. We obtain the reflected electron distribution by excluding particles within a sphere centered on the average background electron momentum with a radius of three times the rms momentum of background electrons. The particle count inside this sphere is subsequently interpolated.
After accounting for uncertainties due to particle statistics and binning procedures, we estimate the uncertainty in both the average and the rms momenta of the background and shock-reflected electrons to be $10^{-3}m_ec$ and $10^{-4}m_ec$, respectively.

The current simulations do not reveal which aspect of the pre-existing turbulence is responsible for the warmer electron distribution in the foreshock, scattering by the additional magnetic fluctuations or enhanced shock corrugation caused by the turbulence. We observe an increased temperature of reflected electrons over the entire foreshock, which suggests that the reflected electrons start with a higher temperature and that the shock corrugation imposed by density fluctuations is an important factor. The steep decline of the parallel momentum in the presence of pre-existing turbulence suggests additional scattering by the magnetic fluctuations, and so both magnetic and kinetic perturbations may be relevant.

Figure~\ref{fig:momentum_profile_bkg} compares the parallel component of the average momentum, $p_{\text{av},\parallel}$, between the homogeneous and turbulent simulations. In the far-upstream region at$x\approx600\lambda_\text{si}$, the parallel momentum is approximately zero for both cases, as expected. Approaching the shock front, it becomes negative, indicating that a return current is established that neutralizes the current of the shock-reflected electrons. The parallel momentum plateaus at around $x\lesssim 400\lambda_\text{si}$ for run H and at $x\lesssim 300\lambda_\text{si}$ for run T.

Figure~\ref{fig:momentum_profile_ref} presents the number density (panel a), the parallel component of the average momentum (panel b), the perpendicular component of the rms momentum (panel c), and the kinetic-energy density of the reflected electrons. The rate at which the number density increases from the upstream region to the shock front (right to left) is similar for the homogeneous and turbulent simulations, but, due to the shorter foreshock in run T, the red dashed line is shifted left by roughly $x\approx100\lambda_\text{si}$, translating to a lower density at a given location. A similar trend is seen in panel b) of the figure.
Panel c illustrates that the rms momentum, $p_{\text{rms},\perp}$, is significantly larger with pre-existing turbulence. As the $p_\perp$ component is Lorentz-invariant, its rms value reflects the perpendicular temperature of the beam of reflected electrons that is higher in the turbulent simulation, at least for a large part of the foreshock. 
The energy density of the reflected electrons within $x\approx100\lambda_\text{si}$ of the shock front is comparable in both simulations, and further out there is much more energy density in the foreshock for homogeneous initial conditions (Panel d).

\begin{figure}[h!]
    \centering
    \includegraphics[width=0.9\linewidth]{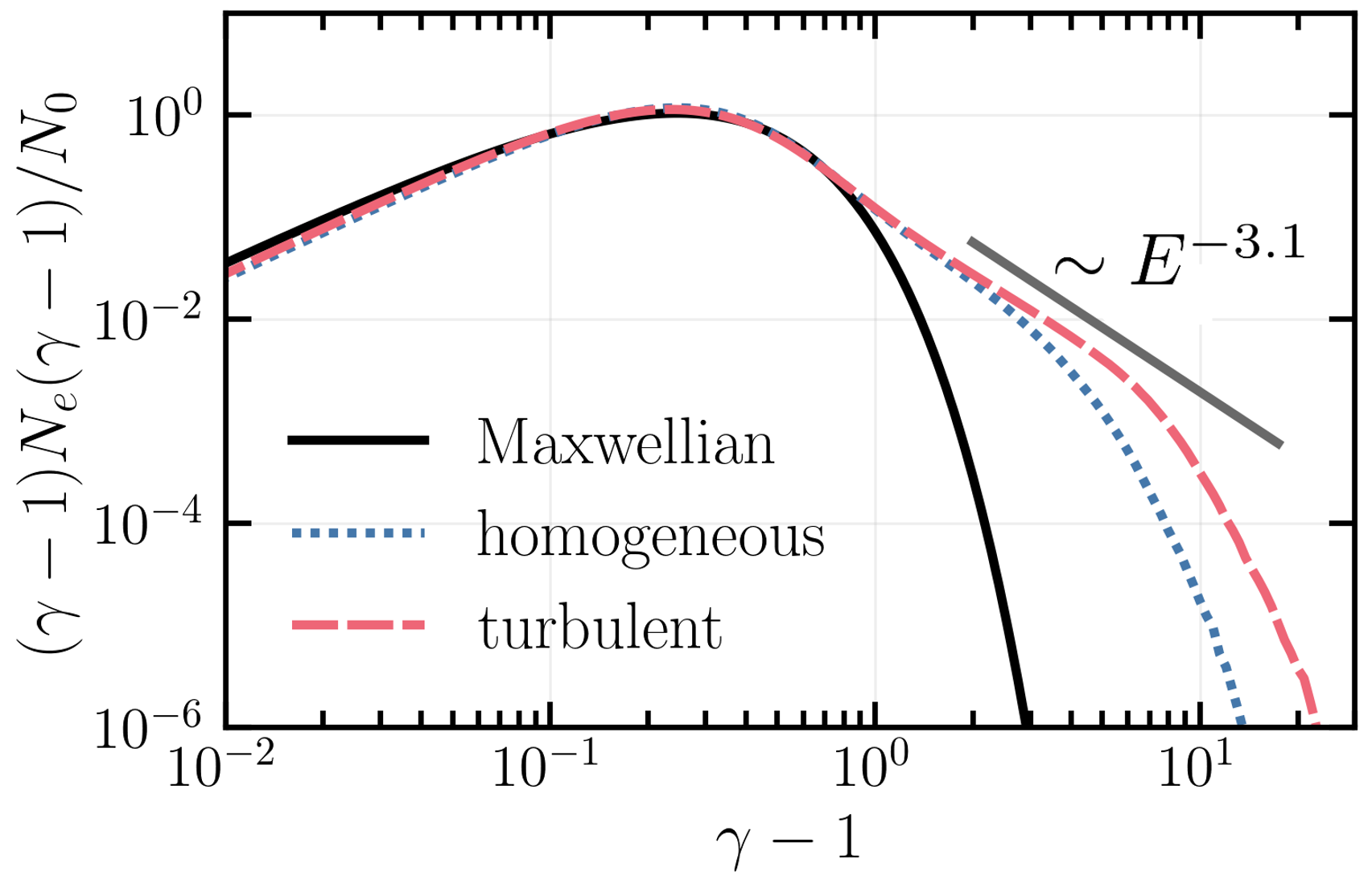}
    \caption{The electron energy spectrum in the downstream regions for runs $\text{T}$ (turbulent; the red-dashed line) and $\text{H}$ (homogeneous; the blue-dotted line). The black solid line denotes a fitted relativistic Maxwellian to the low-energy part of the distribution. The grey solid line represents the slope of the nonthermal tail.}
    \label{fig:oblique_spectrum}
\end{figure}

\subsection{Electron acceleration}

Figure~\ref{fig:oblique_spectrum} shows the electron energy distribution in the downstream region for the simulations $\text{H}$ and $\text{T}$. The spectra are computed in the local plasma rest frame in the region between $20\lambda_\text{si}$ and $90\lambda_\text{si}$ behind the shock at the final time step of our simulations, $t\approx20\Omega_{i}^{-1}$. Low-energy electrons are well represented by a relativistic Maxwellian distribution. As expected from previous studies, both simulations demonstrate a nonthermal spectral tail of the form $N(\gamma)\propto (\gamma -1)^s$ that extends to higher energies for run $\text{T}$. In the homogeneous simulation, the fraction of non-thermal electrons, defined as particles with $\gamma -1 \geq 3$, is approximately equal to 0.14\%. These particles account for about 1.9\% of the total electron energy. For the simulation with pre-existing turbulence, these values are 0.23\% and 3.6\%, respectively, and the maximum Lorentz factor is 40\% higher than without pre-existing turbulence.

\section{Summary} \label{sec:summary}
We have performed 2D3V PIC simulations of nonrelativistic oblique shocks propagating in an electron-ion plasma with pre-existing density fluctuations of amplitude 15\%. The physical parameters are chosen to model conditions in SNRs, and so the shocks have high Mach numbers and propagate in a medium with a plasma beta near unity. Our main objective was to investigate the influence of turbulence on the foreshock properties, the whistler instability, and electron acceleration. The fluctuation level, $\var{n}/n=15\%$, is chosen to avoid excessive heating and is consistent with in-situ measurements of density fluctuations of the local interstellar medium. We evolve the turbulent plasma before we insert it in the shock simulation, at which time the density fluctuations are accompanied by bulk-velocity and magnetic-field fluctuations, the latter typically at the 10-\% level.

We used a sufficiently large simulation box in the transverse direction to capture large nonlinear structures driven by the electromagnetic whistler instability. We directly compared the turbulence's influence on shock properties and electron acceleration to a simulation with a homogeneous upstream medium. Furthermore, we examined the global shock structure along with the magnetic-field amplification. Our key conclusions are:

\begin{enumerate}
   \item The shock structure is slightly modified by the pre-existing turbulence; the overshoot is wider, and the undershoot is less pronounced. In the out-of-plane configuration that we employed, the Weibel instability in the shock ramp is weakened, and so the magnetic field is amplified only by whistler waves ahead of the shock front. We measure amplitudes of the magnetic field about 25\% higher for the simulation with pre-existing turbulence in this region.
    \item A return current is established by background plasma that neutralizes the current of the shock-reflected electrons. This current is caused by background electrons that gradually accelerate as they approach the shock front.
    \item Our simulations indicate that pre-existing density fluctuations cause the length of the electron foreshock to decrease by approximately \text{$\sim 5$} ion gyroradii.
    Furthermore, the shock-reflected electrons have a higher temperature. 
    \item Both simulations show modes with right circular polarization in the upstream rest frame, within the foreshock region. The growth rates     are comparable to the peak growth rate 
    of the oblique whistler instability (for similar plasma parameters). Therefore, we conclude that the observed modes are oblique whistler waves.
    \item The growth rate of the whistler waves is lower by roughly one third in the presence of pre-existing density fluctuations. 
    This is likely due to larger temperature of the shock-reflected electrons, or smaller temperature anisotropy of the beam. For both simulations, the growth rates decrease with time, but the whistler waves start to grow at larger distances from the shock, and their amplitude near it remains the same.
    \item Nonlinear structures similar to those driven by whistlers at the homogeneous shock appear two ion cyclotron times earlier for a run with turbulence, and their maximum size is larger: $7\lambda_\text{si}$ compared to $3\lambda_\text{si}$. 
    \item The nonlinear structures feature high electron pressure due to the confinement of the shock-reflected electrons. 
    This leads to pressure imbalance between the interior and exterior of these structures, causing them to expand as they propagate towards the shock front. Since the nonlinear structures are larger in size and have stronger electromagnetic fields, they eventually distort the shock front to a greater extent when they merge with it.
    \item The high-energy tail of the downstream electron spectrum is well described by $N(\gamma)\propto (\gamma -1)^{-3.1}$ in both cases, but electrons are accelerated more efficiently in the presence of pre-existing turbulence, as both their number (0.23\text{$\%$} vs 0.14\text{$\%$} of total electrons present in the downstream region) at the end of the simulation) and energy (3.6\text{$\%$} vs 1.9\text{$\%$}) densities are higher than for a homogeneous upstream medium. Likewise, the maximum electron energy is higher by 40\%. By scattering on the pre-existing magnetic fluctuations and other waves they spawn in the shock ramp, magnetic fluctuations play an important role in stochastic shock-drift acceleration, and it would be natural to surmise that enhanced turbulence levels are beneficial for the process.
\end{enumerate}

The simulations presented in this study are of higher resolution but have shorter duration ($t\Omega_i  \approx 20$ vs. $t\Omega_i\approx50$) than previous 2D3V PIC simulations of high-Mach-number oblique shock that also employed different electron-ion mass ratios  \citep{Bohdan2022,Morris2023}. 
The previous studies showed that after approximately 30 ion Larmor times the nonlinear cavities reached a size of a few ion skin lengths. In the run with pre-existing turbulence we observe that the nonlinear structures are larger and appear earlier. But the final size of these structures remains uncertain, and they may eventually reach a similar size with and without pre-existing turbulence that may be related to scales of the whistlers at their saturation, the characteristic scales of the pre-existing turbulence, or the simulation-box size.

{The linear growth of the electron whistler instability is independent of the ion-to-electron mass ratio, {but the spatial size of the region where the whistler growth is observed decreases (in ion scales) as the mass ratio increases.} 
{The argument of the exponential in Equation~\ref{eq:growth_rate} can be written as $-2(\gamma/\Omega_e) \cdot M_A^{-1} \cdot m_i/m_e \cdot (x-x_\text{sh})/\lambda_\text{si}$. For the same magnetic-fluctuation amplitude near the shock front, growth rate, and Alfve\'nic Mach number,} the spatial size of the whistler growth region scales inversely with the mass ratio. {As $M_A \lambda_\text{si}$ is the ion gyroradius for the shock speed, which is a proxy of the shock thickness, for a realistic $m_i/m_e$ the spatial scale of wave growth becomes comparable to the shock thickness. It is unclear whether that leads to larger wave amplitudes or to a faster return of reflected electrons to the shock, but in any case we expect an impact on} the electron acceleration efficiency.}

The nonlinear cavities in our simulations exhibit similarities with microscale coherent structures that are commonly observed in turbulent plasmas throughout the heliosphere. High-resolution observations from the Magnetospheric Multiscale (MMS) mission identified kinetic-scale magnetic holes in the Earth's magnetosheath, ranging in size from one ion gyroradius to a few electron gyroradii \citep[see, e.g.,][for a review]{Shi2024}. These structures feature a magnetic field depression, typically accompanied by an increase in the particle density and temperature. One of their possible generation mechanisms is associated with electron temperature anisotropy regions in plasma turbulence that are unstable to oblique electron whistlers that convert to Bernstein modes and then collapse \citep{Espinoza-Troni_2025}. In simulations of decaying turbulence, they seem to emerge from electron velocity shear \citep{2023ApJ...958...11A}. Earth's foreshock transients are phenomena showing a decrease in the magnetic field. Some of them, such as hot flow anomalies, spontaneous hot flow anomalies, and density holes, are associated with depletions in the number density and particle heating in the core, as well as compressions at the edges \citep[see, e.g.,][]{Zhang2020,Zhang2022}. The beam of reflected electrons in our simulations provides substantially different initial conditions than those found in the turbulent plasma of the solar wind or magnetosheath plasma (for example fast relative streaming of two populations of electrons rather than a temperature anisotropy). The fact that the structures we see are not identical yet share similitude with those in heliosphere suggests that such solitary structures are produced in a variety of environments and for range of conditions.

Fluctuations at kinetic scales in the upstream medium of SNRs may originate from an energy cascade starting at considerably larger scales through various fluid instabilities. The wide range of these scales makes it computationally infeasible to follow the self-consistent evolution of plasma turbulence down to kinetic scales, and to study its influence on the shock microphysics and particle acceleration. Global properties of shock evolution, including the effects of large-scale turbulence, can be modelled using fluid approaches. However, they lack information about particle dynamics. Test-particle trajectories \citep{Guo2010,Guo2012b} can only involve high-energy particles whose Larmor radius is resolved by the fluid simulation. Hybrid kinetic simulations, that treat electrons as a massless adiabatic fluid while following ions with the standard PIC method, can probe larger spatial and longer temporal scales than is possible with full PIC, at the expense of neglecting electron kinetic effects. Recent results from such hybrid simulations show significant impact of upstream turbulence on proton acceleration and particle transport at shocks \citep{Trotta2021,Nakanotani2022,Trotta2023}.

Currently, fully kinetic 3D simulations are overwhelmingly challenging. They are limited to lower resolutions and smaller box sizes, and they reach the earliest stages of the system's evolution when the electron foreshock has not yet formed \citep{Matsumoto2017}. \citet{2025PhPl...32e2901O} argue that for our choice of parameters, 2D simulation should provide a reasonable description of ion acceleration, but including all dimensions may affect the evolution and nonlinear development of electron acceleration and instabilities driven at the shock. For example, in the out-of-plane configuration used in this work, the Weibel instability is suppressed, so the interplay between pre-existing modes, whistlers, and the Weibel instability should be in three spatial dimensions.

The preexisting turbulence we model with reduced dimensionality may also exhibit different properties in three dimensions, affecting wave-particle interactions. Kinetic simulations of strong and intermittent Alfvénic turbulence, characterized by large fluctuation amplitudes ($\delta B/B_0\sim 1$), find that the non-linear behaviour of large 3D turbulent plasma is not similar (e.g, larger compressibility, velocity fluctuations strongly coupled to magnetic fluctuations) yet remains consistent with 2D approaches using high resolution kinetic simulations (e.g, dissipation and heating, spectral break at ion scales and subsequent steepening, width of current sheets) \citep{Wan2015,Franci2018,Roytershteyn2019}. \citet{Gary2020} argues that when the background magnetic field is neither strictly perpendicular nor parallel to the simulation plane, both nonlinear processes and microinstabilities might be well represented. The latter are important for turbulence with small-amplitude fluctuations, $\delta B/B_0 \ll 1$, considered in our work. We note that the aforementioned studies examine incompressible turbulence, and only recently compressible turbulence has been investigated with kinetic simulations. Initial findings suggest that the fast-mode cascade might be adequately captured using two dimensions \citep{Hou2025}.

The level of pre-existing compressive turbulence that can be achieved on kinetic scales is limited by particle heating. To maintain sufficiently strong sonic Mach numbers, $M_s \gtrsim 30$, the maximum level of the density fluctuations should be on the order of $\delta n/n \sim 15\%$, with a few times lower amplitudes of the magnetic field fluctuations \citep{Fulat2023}. Future studies will explore the effect of Alfve\'nic turbulence, in which the magnetic-field fluctuations dominate and may reach higher amplitudes without excessive heating.

\begin{acknowledgements}
K.F. acknowledges support by the Simons Foundation as part of Simons Collaboration on Extreme Electrodynamics of Compact Sources (SCEECS). E.M. and M.P. acknowledge support by DFG through grant PO 1508/11-1. T.A. and M.P. acknowledge support for bilateral exchange by DAAD (PPP Projekt 576634589) and JSP. This research was supported by the International Space Science Institute (ISSI) in Bern, through ISSI International Team project \#520
\textit{Energy Partition across collisionless shocks} and by the Munich Institute for Astro-, Particle and BioPhysics (MIAPbP), which is funded by the Deutsche Forschungsgemeinschaft (DFG, German Research Foundation) under Germany´s Excellence Strategy – EXC-2094 – 390783311. 
M.T. acknowledges support by the Czech Science Foundation through the project grant GACR 25-18493s : \textit{Key to Cosmic Rays: A novel approach to a classical problem}.
The authors gratefully acknowledge the computing time made available to them on the high-performance computer ``Lise'' at the NHR Center NHR@ZIB. This center is jointly supported by the Federal Ministry of Education and Research and the state governments participating in the NHR (www.nhr-verein.de/unsere-partner). 
\end{acknowledgements}

\appendix
\section{Polarization of oblique waves} \label{sec:A1} 
We define the polarization of waves by the fluctuating magnetic fields. 
With $\omega$ and $\vb{k}$ representing the wave frequency and the wavevector, a  sum of two 
linearly polarized waves,
\begin{equation}
    \vb{B}(\vb{r},t) = B_0e^{\imath (\vb{k}\cdot\vb{r}-\omega t)}\qty[\vu{e}_1 \mp \imath\, \vu{e}_2],
\end{equation}
where $\vu{e}_1$ and $\vu{e}_2$ are unit vectors that, together with $\vu{k}$, form a right-handed orthogonal set of unit vectors. The ``+'' sign in the above expression denotes a left-hand circularly polarized wave, while the ``-'' sign denotes a right-hand circularly polarized wave.

For a wave propagating along the $x$-axis in Cartesian coordinates, the orthogonal set of unit vectors $(\vu{e}_1,\vu{e}_2,\vu{k})$ can be $(-\vu{z},\vu{y},\vu{x})$. Since the wave is transverse, the magnetic-field vector has $y$ and $z$ components. Our simulations are performed in the $xy$ plane, in which waves can have an arbitrary orientation $\vb{k}=(k_x,k_y)$. Rotating the coordinate system by the angle between the wavevector and the $x$-axis, $\theta(\vb{k})=\angle (\vb{k},\vu{x}) = \arctan(k_y/k_x)$, gives
\begin{equation}
    (\vu{e}_1,\vu{e}_2,\vu{k})=(-\vu{z},-\sin\theta\vu{x}+\cos\theta\vu{y},\vu{k}).
\end{equation}
The magnetic vector of a circularly polarized wave,
\begin{equation}
    \vb{B}(\vb{r},t) = B_0e^{\imath (\vb{k}\cdot\vb{r}-\omega t)}\qty[\pm \imath \sin\theta\vu{x} \mp \imath \cos\theta\vu{y}-\vu{z}] = B_x\vu{x}+B_y\vu{y}+B_z\vu{z},      
\end{equation}
can then be split into amplitudes for the right and left hand polarization,
\begin{align}
    B_R = B_z-\imath \qty(B_x\sin\theta-B_y\cos\theta) \label{eq:B_R}, \\
    B_L = B_z+\imath \qty(B_x\sin\theta-B_y\cos\theta) \label{eq:B_L}.
\end{align}
For the right (left) hand polarization, $B_R$ ($B_L$) equals $-2B_0e^{\imath (\vb{k}\cdot\vb{r}-\omega t)}$, while zero for the left (right) hand polarization.

\section{Polarization measurements in pre-existing turbulence} \label{sec:A2}
\begin{figure}[h!]
    \centering
    \includegraphics[width=0.4\linewidth]{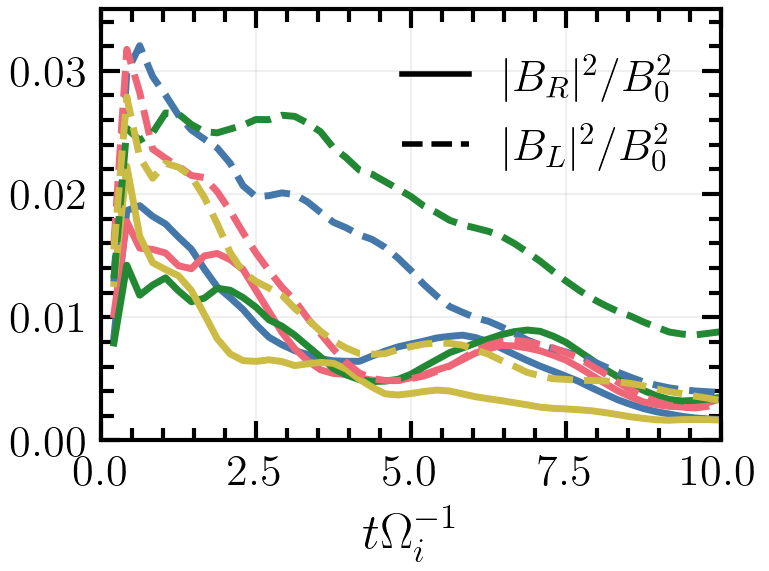}
    \caption{The evolution of the energy density of the right and left circularly polarized waves for different plasma slabs (depicted with different colours here for each slab). The energy density is normalized to the energy density of the initial background magnetic field.}
    \label{fig:polar_slabs}
\end{figure}

In our shock simulations, the upstream medium consists of slabs of compressive turbulence, which are simulated separately in periodic box simulations and then injected into a shock simulation. To investigate the properties of wave modes in the electron foreshock, particularly their polarization, we first examine them in the pre-defined slabs of turbulent plasma. Figure~\ref{fig:polar_slabs} shows the evolution of the energy density in the right and left circularly polarized modes for four example plasma slabs, marked by different colors. The energy density is calculated using Equations~\ref{eq:FT_B_R} and \ref{eq:FT_B_L}. We typically inject slabs after the initial rapid period of evolution, at about $\Omega_i^{-1} t \approx2$. At this stage $B_L$ is stronger than $B_R$, but as the turbulence evolves, their magnitudes become comparable. To summarize, in the upstream region with pre-existing turbulence we expect to have $B_R^2 \sim B_L^2 \sim 0.01 B_0^2$, which is consistent with the values in Figure~\ref{fig:polar_shock}.

\bibliography{references}{}
\bibliographystyle{aasjournal}

\end{document}